\documentclass[aps,floats,12pt]{revtex4}
\input epsf.tex
\epsfclipon

\newcommand{\NC}{\newcommand}
\NC{\beq}{\begin{equation}}
\NC{\eeq}{\end{equation}}
\NC{\beqa}{\begin{eqnarray}}
\NC{\eeqa}{\end{eqnarray}}
\NC{\lra}{\leftrightarrow}

\NC{\sss}{\scriptscriptstyle}
\NC{\lsim}{\mbox{\raisebox{-.6ex}{~$\stackrel{<}{\sim}$~}}}
\NC{\gsim}{\mbox{\raisebox{-.6ex}{~$\stackrel{>}{\sim}$~}}}

\begin{document}

\title{Codimension-two branes in six-dimensional supergravity\\ and 
the cosmological constant problem}
\author{J\'er\'emie Vinet and James M. Cline}
\affiliation{\\
Physics Department, McGill 
University,\\3600 University Street, Montr\'eal, Qu\'ebec, Canada H3A 2T8}
\email{jcline@physics.mcgill.ca,vinetj@physics.mcgill.ca}

\date{\today}

\begin{abstract}
{We investigate in detail recent suggestions that codimension-two braneworlds in 
six dimensional supergravity might circumvent Weinberg's no-go theorem for self-tuning
of the cosmological constant.  The branes are given finite thickness 
in order to 
regularize mild singularities in their vicinity, and we allow them to have
an arbitrary equation of state.  We study perturbatively the time evolution of the solutions
by solving the equations of motion linearized around a static 
background.
Even allowing for the most general possibility of warping 
and nonconical singularities, the geometry does not relax to a 
static solution when the brane stress-energies are perturbed.  Rather, 
both the internal and external geometries become time-dependent, and
the system does not exhibit any self-tuning behavior.}
\end{abstract}


\maketitle

\section{Introduction}

There has been considerable recent interest in the possibility that braneworld
constructions give new possibilities for solving the cosmological constant
problem \cite{Arkani-Hamed:2000eg,Kachru:2000hf,Forste:2000ft,Chen:2000at,Carroll,Navarro1,
Gunther:2003zn,Cline:2003ak,Aghababaie:2003wz,Navarro2,Kehagias,VC}.  
One rationale is that intrinsically extra-dimensional effects might 
provide a loophole to Weinberg's no-go theorem against self-tuning mechanisms, which is formulated in 4D \cite{Weinberg}.
Such attempts first arose using 5D models, where solutions could
be constructed with the property that, regardless of the 4D cosmological
constant (brane tension), the universe was static \cite{Arkani-Hamed:2000eg,Kachru:2000hf,Forste:2000ft}.  Unfortunately it was shown
that these solutions had singularities in the bulk which when regularized 
corresponded to the presence of an additional fine-tuned brane; moreover 
these solutions had flat directions which would give rise to nonstatic solutions
when the branes were perturbed away from their fine-tuned values.  Other 
attempts involving black holes in the bulk were also plagued by instabilities \cite{Binetruy:2000wn,Csaki:2000dm,Csaki:2001mn,Cline:2001yt}.

The case of a 6D bulk with (codimension-two) 3-branes has proved to be 
somewhat more subtle \cite{Chen:2000at,Carroll,Navarro1,
Gunther:2003zn,Cline:2003ak,Aghababaie:2003wz,Navarro2,Kehagias,VC,Sundrum:1998ns,GS,Nihei:2000gb,Ponton:2000gi,Giovannini:2001hh,Kanti:2001vb,Neupane:2001kd,chacko,KMPR,Carroll:2001ih,BCCF,Corradini:2002ta,GGP,Aghababaie:2003ar,Nilles:2003km,Lee:2003wg,deCarlos:2003nq,Gregory2,Burgess:2004kd,Graesser:2004xv,Burgess:2004yq,kannosoda,Wu:2004gp,GP,
Burgess:2004xk,Burgess:2004dh,Burgess:2004ib,navarrosantiago,Redi,Kofinas:2004ae}.
The conical singularity induced in the 6D Ricci scalar
by the codimension-two source exactly cancels the contribution of the
delta-function source itself when the 6D action is dimensionally reduced.  
The contribution of the brane tension to the 4D cosmological constant thus
disappears automatically; the 4D effective action does not see it---it is as
though it has been fine-tuned out of the theory, but in fact it was the
dynamics of 6D general relativity which canceled it.    Superficially, this
looks exactly like  the kind of inherently extra-dimensional effect which
might well circumvent Weinberg's theorem.  In practice however, the idea has
not yet been shown to work.  The problem is that the tension through its
gravitational couplings affects other terms in the effective action besides
the bare delta function source.  In the models that have been studied so far,
the sensitivity of these extra terms to the brane tension
spoils the hoped-for self-tuning effect that would manifest
itself when the tension undergoes a dynamical change, as during a phase
transition.

The failure of self-tuning in the simplest model, where the extra dimensions
are compactified by the flux of a U(1) gauge field, was demonstrated in 
references \cite{VC} and \cite{GP}.  In \cite{GP}, it was also argued that the
same observations rule out self-tuning in supergravity versions of the model.
We disagree with this point of view, as explained below.  Conflicting
arguments in the literature hold out the hope that self-tuning could still
work in 6D SUGRA models \cite{Burgess:2004dh,Burgess:2004ib}, so it is worthwhile to investigate in
detail, given the notorious intractability of the cosmological constant
problem.  

We review the formalism of codimension-two braneworld models in section II,
with and without supersymmetry, and the details of arguments for and against
self-tuning which have been made for the SUSY case.  We also explain why it is
useful to consider finite-thickness branes.  In section III we set up and
solve the leading-order equations of motion  for the SUSY case, treating  the
time-dependent brane sources as perturbations.  Section IV focuses on a
particular class of warped background solutions which is crucial to settling
the self-tuning issue; we demonstrate that these solutions are adequately
described by our perturbative treatment so that self-tuning, if present, would
be visible within our calculational framework.  In Section V we impose the
boundary conditions at the branes, which reduces the general solution of
section III to the particular ones of cosmological interest, in the case of
time-dependent change of brane tension.  We show that the system does not relax
back to a static solution, but rather evolves like a universe with nonvanishing
cosmological constant---there is no self-tuning.  These results are discussed
in section VI, and technical details are given in the appendix.

\section{Review of codimension-two braneworlds}

In this section we will briefly review how self-tuning is known to fail in the
nonsupersymmetric 6D models, and the issues surrounding whether it can work in
SUGRA versions of the models.

\subsection{Einstein-Hilbert gravity}
The simplest model which allows for compactification of the two extra dimensions
has the action
\beqa
\label{EHaction}
S = \int d^{\,6}x \sqrt{-g} \left[\frac{M_6^4}{2}{\cal R} 
-\frac{1}{4}
F_{MN}F^{MN}-\Lambda_6\right] + S_{\rm branes},
\eeqa
where
\beq
	S_{\rm branes} =   -\int d^{\,4}x\, d^{\,2}y\, \sqrt{-g} \, 
	\sum_i \tau_i {\delta^{(2)}(\vec y - \vec y_i)\over \sqrt{g_2}}
\eeq
The interplay between a positive bulk cosmological constant and a gauge
field flux $F_{r\theta}$ gives an energetically preferred size for the
compact space.  (It should be noted that this type of model was initially 
studied in the mid 1980's, before the concept of branes was introduced \cite{Rubakov:1983bz,Randjbar-Daemi:1985wg,Wetterich:1984rv,Wetterich:wd,Gibbons:1986wg}).

Solutions exist with static extra dimensions and which are maximally symmetric
in the large four dimensions,\footnote{solutions with $H^2<0$ can be
analytically continued to 4D AdS} 
\beqa
	ds^2 &=& -dt^2 + e^{2Ht}\, dx^2 + k^{-2}\left(dr^2 + {\sin^2r}\,
	d\theta^2\right)\nonumber\\
	A_\theta &=& {\beta\over k^2}\, (1\pm \cos r)	
\eeqa
with $k^2 = M_6^{-4}(\Lambda_6/2 +  3\beta^2/4)$, 
$H^2 = M_6^{-4}(\Lambda_6/6 - \beta^2/12)$, the coordinates $r$ and 
$\theta$ range over $(0,\pi)$ and $(0,2\pi)$ respectively, and 
where the 6D Ricci scalar has the form
\beq
	R = {2\over M_6^4}\left(\sum_i \tau_i 
	{\delta^{(2)}(\vec y - \vec y_i)\over \sqrt{g_2}}
	+ 6H^2+ k^2 \right)
\eeq
Thus, the singular parts of the action cancel exactly when evaluated for the
classical solutions, and the effective 4D cosmological constant is determined
entirely by the smooth parts:
\beq
\label{lambda4}
	\Lambda_4 = V_2\left(\frac12{\Lambda_6} - \frac14{\beta^2}\right),
\eeq
where $V_2$ is the volume of the compact manifold,
\beq
\label{volume}
	V_2 = 2\,(2\pi - \delta)\, k^{-2},
\eeq
and $\delta$ is the deficit angle associated with conical singularities 
induced by 3-branes, so that in the presence of brane tension, the coordinate $\theta$ 
defined above ranges from $0$ to $2\pi -\delta$.
It requires a fine tuning of the bulk quantities $\Lambda_6$ and $\beta$ to have a vanishing
cosmological constant, so the cancellation of the singular parts would not
in itself explain the smallness of $\Lambda_4$.  Nevertheless it is suggestive;
for example if supersymmetry in the bulk caused (\ref{lambda4}) to vanish, then
SUSY could be broken on the branes without affecting $\Lambda_4$.

However this seemingly elegant idea does not work, as was explicitly shown in
\cite{VC}.  The reason can be understood intuitively as follows \cite{GP}.  The
simplest version of the model has two branes with equal tension $\tau$ at
antipodal points of the spherical extra dimensions, which produces conical
defects and a deficit angle $\delta = \tau/M_6^4$.  The volume of the extra
dimensions depends linearly on the deficit angle, eq.\ (\ref{volume}).  The
flux of the gauge field in the extra dimensions is proportional to the volume,
$\Phi = F_{r\theta} V_2$.  This flux is conserved, so if one tries to change the
brane tensions, the field strength has to  change to compensate for the change
in volume.  Any such change in $F_{r\theta}$ spoils the tuning needed to make
$\Lambda_4$ vanish in  (\ref{lambda4}).  Thus $\Lambda_4$ {\it is} linearly
dependent on the brane tensions---not through the delta function terms in the
original action, but as a consequence of the conical defect which they produce
on the bulk geometry.

\subsection{Supergravity}
A supergravity version of the above model involving chiral fields
$\Phi_a$ and a dilaton $\phi$ can be derived from the action
\beqa
\label{susyaction}
S &=& \int d^{\,6}x\, \sqrt{-g} \left[\frac{M_6^4}{2}\,\left({\cal R} 
-\partial_M\phi\,\partial^M\phi\right) -\frac{1}{4}
e^{-\phi}F_{MN}F^{MN}
-\frac{1}{2}h(\Phi)_{ab}\,\partial_M\Phi^a\partial^M\Phi^b\right.\nonumber\\
&&\left.-e^{\phi}\,v(\Phi)\phantom{\frac{M_6^4}{2}\!\!\!\!\!\!\!\!\!}\right] + S_{\rm branes},
\eeqa
One can consistently assume that the potential $v(\Phi)$ is minimized when
the fields $\Phi_a=0$, and thus ignore these fields and treat $v(\Phi)$
as the bulk cosmological constant, $\Lambda_6$.

Comparing with the nonSUSY model, it is easy to see that there exist solutions
with constant $\phi$, where everywhere we have replaced $\Lambda_6\to\Lambda_6
e^\phi$ and $\beta^2\to\beta^2 e^{-\phi}$, provided that $\phi$ takes the
value $\phi_0$ such that its induced potential is stationary,
\beq
	V'(\phi_0) = -\frac12\beta^2\, e^{-\phi_0} + \Lambda_6 \, e^{\phi_0}
\equiv 0
\eeq
Comparing to (\ref{lambda4}), this is precisely the value needed to make the effective cosmological
constant vanish,
\beq
\label{lambda4a}
\Lambda_4 \to V_2\left(\frac12{\Lambda_6}\, e^{\phi_0} 
- \frac14{\beta^2}\, e^{-\phi_0} \right) = 0	
\eeq
This seemingly automatic adjustment of $\Lambda_4$ to zero is one reason 
that previous authors \cite{Aghababaie:2003wz,Aghababaie:2003ar,Burgess:2004kd,Burgess:2004xk,Burgess:2004dh,Burgess:2004ib} have been encouraged to seek a
self-tuning mechanism in this model.  

More generally, there are three classes of solutions for this model whose
4D metric is maximally symmetric \cite{GGP}, and all of them have vanishing
vacuum energy.  The solution displayed above is the simplest one, which
corresponds to putting branes of equal tension at the antipodes of the the
compact space.  A second class of solutions has branes of  unequal tension, in
which case the metric no longer factorizes, but instead is warped.  The third
class of solutions, which are also warped, is more  exotic:  it has nonconical
singularities at the antipodal points whose stress energy does not correspond
to simple brane sources.  In particular, the $T_{rr}$ and $T_{\theta\theta}$
stress tensor components are nonvanishing for these backgrounds.

\subsubsection{The argument against self-tuning}
In ref.\ \cite{GP}, it was argued that, rather than displaying self-tuning,
the SUGRA model is a clear example of Weinberg's no-go theorem.  Ref.\
\cite{GP} dimensionally reduces the model with two equal-tension branes
to 4D gravity coupled to 
the dilaton $\phi$ and the radion $\psi$, which enters
the metric ansatz in the form
\beq
\label{GPansatz}
	ds^2 = g_{\mu\nu}\, dx^\mu\, dx^\nu + M_6^{-2}e^{-2\psi} (dr^2 + \sin^2 r\,
d\theta^2)
\eeq
The potential which results for the two scalar fields is found to have the form
\beq
\label{GPpot}
	V(\psi,\phi) = M_6^{-4}\,e^{\sigma_2}\,\left({\beta^2\over
2\alpha^2}\,e^{-2\sigma_1} - 2M_6^{6}\,e^{-\sigma_1} + 2\Lambda_6\right)
\eeq
where $\sigma_2 =2\psi-\phi$, $\sigma_1 = 2\psi+\phi$, and $\alpha$ is related
to the deficit angle induced by the branes, $\alpha = 1 - \tau/(2\pi M_6^4)$.
This potential looks different from the dilaton potential in (\ref{lambda4a})
because in the latter, the Einstein equations have already been imposed, which
fixes $\psi$ in terms of $\beta^2$ and $\Lambda_6$, whereas this has not been
done in (\ref{GPpot}).  Ref.\ \cite{GP} goes on to observe that, starting from
some configuration in which $V(\psi,\phi)=0$, any change in the brane tensions,
hence in the value of $\alpha$, will make $V(\psi,\phi)$ nonzero, and induce
a runaway potential for the field $\sigma_2$.  

If the 4D potential (\ref{GPpot}) really captures the essential features
of the model, then clearly there is no self-tuning; however this ansatz
describes only the unwarped solutions.  If self-tuning occurs, it might
be through a transition from an unwarped to a warped solution, especially
an exotic nonconically singular one, which cannot be described by the
ansatz (\ref{GPansatz}).  

More generally, the proposal of \cite{Aghababaie:2003wz,Aghababaie:2003ar,Burgess:2004kd,Burgess:2004xk,Burgess:2004dh,Burgess:2004ib} involves large extra
dimensions, with $M_6^{-1}\sim 0.1$ mm, where the Kaluza-Klein mass gap is of
the same order of magnitude as the mass of the stable direction $\sigma_1$ in
the potential (\ref{GPpot}).  The argument of \cite{GP}  has thus been called
into question, on the basis that there is no justification for ignoring the
role of light KK modes while keeping other modes that are of the same energy in
the effective 4D description.  This objection is similar in spirit to the
previous one, in that mild warping of the background metric can be thought of
as a coherent state of KK modes.

\subsubsection{The argument for self-tuning}

One reason for suspecting a self-tuning mechanism in the Salam-Sezgin
model is the observation by \cite{GGP} that the only solutions with 
maximal symmetry in the 4D part of the metric are those with vanishing
curvature in 4D, that is, Minkowski space.  These solutions fall into
two classes, one of which has only conical singularities (or possibly no
singularities).  It has the warped line element and dilaton profile
\beqa
	ds^2 &=&  \sqrt{\hat r^2 + \hat r_1^2\over \hat r^2 + \hat r_0^2} \left( dx^2 + 
	{d\hat r^2\over (1 + \hat r^2/\hat r_0^2)^2} +  {\hat r^2 d\theta^2 \over (1 + \hat r^2/\hat r_1^2)^2} 
	\right)\nonumber\\
	\phi(\hat r)  &=&  \phi_0+  \ln\left( {\hat r^2 + \hat r_1^2\over \hat r^2 + \hat r_0^2} \right) 
\eeqa
where 
\beq
	\hat r_0^2 = {4 e^{-\phi_0} \over \Lambda_6};\quad \hat r_1^2 = {8 e^{\phi_0}\over \beta^2}
\eeq

In the special case where $\hat r_0^2 = \hat r_1^2$ ({\it i.e.,} $\Lambda_6 e^{\phi_0} =\beta^2 e^{-\phi_0}/2$), we recover
the unwarped spherical football-shaped solutions, which have branes of equal
tension at the antipodal points.  For $\hat r_0^2 \ne \hat r_1^2$, the tensions of the
branes are related to each other by
\beq
	\left(1 - 4\tau_+/M_6^4\right)\left(1-4\tau_-/M_6^4\right) = \hat r_1^1/\hat r_0^2 = N^2
\eeq
where $N$ is an integer, related to the quantization of gauge field flux on the
compact manifold.

The most general solutions are more complicated; they are given by
\beqa
	ds^2 &=& W^2 \, dx^2 + a^2\left( W^8\, d\eta^2 +  d\theta^2\right)\nonumber\\
	\phi(\eta) &=& \phi_0 + 4\ln W + 2\lambda_3\eta
\eeqa
with
\beqa
W^4 &=& {\lambda_2\over\lambda_1}\, \sqrt{e^{\phi_0}\beta\over 2\Lambda_6}\,
	{\cosh \lambda_1(\eta-\eta_1)\over \cosh \lambda_2(\eta-\eta_2)}
\nonumber\\
a^{-4} &=&	\sqrt{\Lambda_6\over 8} \beta^3\lambda_1^{-3}\lambda_2^{-1}\,
	e^{-2\lambda_3\eta}\, \cosh^3 \lambda_1(\eta-\eta_1)\,
	 \cosh \lambda_2(\eta-\eta_2)
\eeqa
and $\lambda_3 = \sqrt{\lambda_2^2-\lambda_1^2}$.  When $\lambda_3=0$, this
reduces to the special solutions above.

One can see the nonconical nature of the singularities when $\lambda_3
\neq 0$ by performing a coordinate redefinition \cite{Burgess:2004dh}.  
Writing $dr = aW^4 d\eta$, in the limit $\eta \rightarrow \pm \infty$
\beqa
ds_2^2 = dr^2 + r^{2\alpha_{\pm}} d\theta^2
\eeqa
where
\beqa
\alpha_{\pm} = \frac{\pm 2\lambda_3+\lambda_2+3\lambda_1}{\pm 2\lambda_3+5\lambda_2-\lambda_1}.
\eeqa
As soon as $\lambda_3\neq 0$, one finds that $\alpha_{\pm}\neq 1$ and the metric near the 
branes no longer corresponds to that of a simple conical singularity.  Rather, we 
find a geometry more like that of a trumpet, where the circumference of closed circles around the 
singular points is not linearly proportional to their radius.  It is the 
existence of such solutions which has given hope of circumventing the argument 
of \cite{GP}.  Indeed it is not excluded that under a perturbation 
to the branes' tensions, the system would relax to a static solution 
with nonzero $\lambda_3$ rather than move to a runaway solution with purely 
conical singularities.  It is the presence of such behaviour that we will be 
looking to confirm or negate in the analysis that follows.

\subsection{The need for thick branes}
The solutions we have summarized aboved only concern pure tension branes and 
are therefore not suited for the analysis we wish to carry out, where the 
energy density on the branes is time dependent, and its equation of state 
must be different from $\rho = -p$.  In order to study this more 
general case, it is necessary to regularize the singularity at the brane 
positions by giving the branes finite thickness.  The reasons for this were 
explained in detail in \cite{VC}.  We will only recall here the main line of 
reasoning.  

For a metric of the form 
\beqa
ds^2 = -n^2(\bar r,t)dt^2 + a^2(\bar r,t)d\vec x^2 + f^2(\bar r,t)(d\bar r^2+\bar r^2 d\theta^2)
\eeqa
codimension-two branes will appear in the stress energy tensor $T^A_B$ as 
$(\rho,p)\times \delta(\bar r)/[2\pi f^2(\bar r,t) \bar r]$.  There must be terms in the 
Einstein tensor that provide the necessary delta  functions to match the ones 
from the stress-energy tensor.  Since the two dimensional delta function is 
given by
\beqa
\delta(\bar r)  = 2\pi \bar r\nabla^2\ln(\bar r)
\eeqa
the delta function terms in the Einstein tensor will come from terms like 
$\nabla^2 \ln(n)$, $\nabla^2 \ln(a)$ and $\nabla^2 \ln(f)$.  In order to accommodate 
general equations of state, the first two of these must be nonzero near the 
branes, which means that the warp factor must scale like $\bar r^{\alpha(t)}$ as 
$\bar r\rightarrow 0$.  As was done in \cite{VC} to treat the Einstein-Hilbert 
case, we will be smoothing out this mildly singular behaviour by giving the 
brane finite thickness, allowing us to find well-behaved solutions with 
arbitrary equations of state, and thus arbitrary time dependence of the stress-energy 
tensor, on the branes.  

\section{Thick codimension-two braneworld in 6D supergravity}

For the reasons that have just been exposed, we need to regularize the 
branes by giving them nonzero thickness in order to consistently study 
their cosmology in situations where their stress-energy is not simply 
constant.  We will be following the formalism of \cite{VC}, where such a 
construction was used to conclusively rule out the presence of self-tuning 
in the context of six dimensional Einstein-Hilbert gravity.  The only new 
ingredient here will be the presence of a dilaton field.

We start with the most general dynamical metric having axial symmetry in 
the extra dimensions
\beqa
\label{metric}
ds^2 = -n(r,t)^2dt^2+a(r,t)^2d\vec x^2+b(r,t)^2dr^2+c(r,t)^2d\theta^2
+2E(r,t)drdt
\eeqa
and perturb its components in the following manner
\beqa
n(r,t) &=& e^{A_0(r)+N_1(r,t)};\qquad a(r,t) = a_0(t)e^{A_0(r)+A_1(r,t)};
\qquad b(r,t) 
= b_0(t)e^{B_0(r)+B_1(r,t)};\nonumber\\
c(r,t) &=& c_0(t)e^{C_0(r)+C_1(r,t)};\qquad E(r,t) = E_1(r,t).
\eeqa
The zeroth order solutions are for simplicity taken to be ones corresponding
to zero-tension branes, and the perturbations represent the effect of
small time-dependent brane stress-energies.  
We also expand the dilaton and gauge potential as
\beqa
e^{\phi(r,t)} = \varphi_0(t)e^{\phi_0(r)+\phi_1(r,t)};\qquad 
A_{\theta}(r,t) = 
A_{\theta}^{(0)}(r)+A_{\theta}^{(1)}(r,t).
\eeqa
The stress energy tensor is given as
\beqa
T^a_b = t^a_b+s^a_b+{s_*}^a_b
\eeqa
where $t^a_b$ represents contributions from all bulk fields, and $s^a_b$ 
and ${s_*}^a_b$ represent the brane content.  Since we are 
assuming that the branes are only present at the perturbative level,  
their stress energy can be written as
\beqa
s^t_t &=& -\rho\,\theta(r_0-r)-\delta(r_0-r){\cal F}_0(t);\qquad
s^i_i = p\,\theta(r_0-r)-\delta(r_0-r){\cal F}_0(t);\nonumber\\
s^r_r &=& p^r_r\,\theta(r_0-r);\qquad
s^{\theta}_{\theta} = p^{\theta}_{\theta}\,\theta(r_0-r);\qquad
s^r_t= p^r_t\,\theta(r_0-r);\qquad\ 
s^t_r= p^t_r\,\theta(r_0-r);\nonumber\\
{s_*}^t_t &=& -\rho_*\,\theta(r-r_*)-\delta(r-r_*){\cal F}_*(t);\qquad
{s_*}^i_i = p_*\,\theta(r-r_*)-\delta(r-r_*){\cal F}_*(t);\nonumber\\
{s_*}^r_r &=& {p_*}^r_r\,\theta(r-r_*);\qquad
{s_*}^{\theta}_{\theta} = {p_*}^{\theta}_{\theta}\,\theta(r-r_*);\qquad
{s_*}^r_t= {p_*}^r_t\,\theta(r-r_*);\qquad
{s_*}^t_r= {p_*}^t_r\,\theta(r-r_*)\nonumber\\
\eeqa
where $\theta(r)$ are Heaviside step functions.  Our branes are therefore 
represented by cores of finite radii $r_0$ and $r_*$ located around the 
poles of an axially symmetric compact internal space.  For simplicity, 
we will assume $r_*$ and $r_0$ are such that both branes have the 
same thickness at the level of the unperturbed background geometry.  

Notice the inclusion of one-dimensional delta function terms in the 
brane stress-energy.  In \cite{VC}, these came from expanding 
the step function in a Taylor series, treating the time dependence of 
the thickness as a perturbation.  Here, since the background tension is 
assumed to vanish, such terms do not appear unless we put them in by hand.  
The end result however can be shown to be exactly the same: the 
one-dimensional delta functions effectively encode the time dependence of 
the brane thickness.  (See \cite{kannosoda} and \cite{navarrosantiago} for 
discussions on this point).  These functions are not assumed to be the same 
at both branes (i.e. we will not be demanding that ${\cal F}_0 = 
{\cal F}_*$), so that the above assumption that the background thicknesses 
are equal is not overly constraining.

Finally, we will assume that time derivatives 
of ${\cal O}(\rho)$ perturbations are of ${\cal O}(\rho^{3/2})$, which is 
implied by the usual law for conservation of energy 
$\dot \rho \sim (\dot a/a) \rho \sim \rho^{3/2}$.

\subsection{Background solutions}

With the above ansatz for the metric and matter content, working in 
coordinates where $B_0(r)=0$ one can show that 
\beqa
{\partial A^{(0)}_{\theta}\over \partial r} = -\beta e^{-4A_0(r)+C_0(r)+\phi_0(r)}.
\eeqa
Since at this level there is no stress-energy on the brane to induce 
warping or singularities, the only possible solution is the 
perfectly spherical static ``football'' solution \cite{Aghababaie:2003wz}, 
where
\beqa
A_0(r)&=&0\\
\phi_0(r) &=& \phi_0\\
e^{C_0(r)}&=&\frac{\sin(kr)}{k}.
\eeqa
(Note that the coordinate $r$ defined here differs from the one in section IIA by 
the rescaling $r\rightarrow kr$.  In these coordinates, $\theta$ ranges over $(0,2\pi)$ 
even in the presence of a deficit angle, while $r$ of course ranges over $(0,\pi/k)$.)
Substituting these into the background equations of motion, one finds
\beqa
\varphi_0(t)v(\Phi_0) &=& \frac{\beta^2}{2\varphi_0(t)c_0(t)^2b_0(t)^2}\\
\frac{k^2M_6^4}{b_0(t)^2} &=& \varphi_0(t)v(\Phi_0) +\frac{\beta^2}{2\varphi_0(t)c_0(t)^2b_0(t)^2}.
\eeqa
One can readily check that these imply that
\beqa
\beta^2&=&\frac{k^4M_6^8}{2v(\Phi_0)}.
\eeqa
and that the time dependent functions $b_0(t), c_0(t)$ and $\varphi_0(t)$ 
obey
\beqa
b_0(t) &=& c_0(t)\\
\varphi_0(t) &=& \frac{\beta}{\sqrt{2v(\Phi_0)}c_0(t)^2}.
\eeqa
Notice here a key difference with the Einstein-Hilbert case, where there 
is no dilaton.  In this case, we would have $\varphi_0(t)e^{\phi_0}v(\Phi_0)\rightarrow \Lambda_6$ 
and $\beta^2/[\varphi_0(t)e^{\phi_0}]\rightarrow \beta^2$, and we would instead find 
that $b_0(t)$ and $c_0(t)$ must be constants.  This difference can be understood as follows.  In the 
Einstein-Hilbert case, since the radion is 
stabilized we find in our perturbative analysis that the scale factor 
of the internal space is static.  In the SUSY case, there are two degrees of 
freedom: the radion and the dilaton.  Only one linear combination of the two 
is stable, while the other corresponds to a flat direction.  It is this mode 
which appears here as a new dynamical entity.

\subsection{Perturbed equations of motion}

Before writing down the equations of motion to linear order in the 
perturbations, it is useful as was done in \cite{VC} to find a set 
of variables which are invariant under the gauge transformation 
\beqa
t\rightarrow f(r,t);\qquad r\rightarrow g(r,t).
\eeqa
One convenient set is
\beqa
Z &=& N_1' - A_1';\qquad\quad
W = 3A_1'+N_1';\qquad
X = \frac{C_1'}{C_0'}-B_1-\frac{C_0''}{C_0'^2}C_1;\nonumber\\
&& Y = {A_{\theta}^{(1)}}'-{A_{\theta}^{(0)}}'(B_1+C_1);\qquad\qquad
U = \dot A_{\theta}^{(1)}-\frac{{A_{\theta}^{(0)}}'}{C_0'}\dot C_1;
\nonumber\\
\tilde\rho &=& -s^t_t-{s_*}^t_t;\qquad
\tilde p = s^i_i+{s_*}^i_i;\qquad
\tilde p^r_t =s^r_t+{s_*}^r_t;\qquad
\tilde p^t_r = s^t_r+s^r_t+{s_*}^t_r+{s_*}^r_t;\nonumber\\
\tilde p_5 &=& s^r_r+{s_*}^r_r;\qquad
\quad \tilde p_6 = s^{\theta}_{\theta}-s^r_r+{s_*}^{\theta}_{\theta}
-{s_*}^r_r,
\eeqa
in terms of which the perturbed equations of motion become
\beqa
\label{THminRR}
&& W'-C_0'W = c_0(t)^2\frac{\tilde p_6}{M_6^4}\\
\label{TTminXX}
&& \frac{Z'+C_0'Z}{c_0(t)^2} = 2  \left[\frac{\ddot a_0}{a_0}
-\left(\frac{\dot a_0}{a_0}\right)^2-\frac{\dot a_0}{a_0}\frac{\dot c_0}
{c_0}
+2\left(\frac{\dot c_0}{c_0}\right)^2+\frac{\ddot c_0}{c_0}\right]
+\frac{1}{M_6^4}(\tilde\rho +\tilde p)\\
\label{FEOM6}
&& Y'-C_0'Y -e^{\phi_0}\beta e^{C_0(r)}\left( W-\phi_1'\right) = 0\\
\label{RR}
&& \frac{C_0'W}{c_0(t)^2}+\frac{\sqrt{2v(\Phi_0)} e^{-C_0(r)}}{c_0(t)^2M_6^4}Y 
+\beta\frac{\sqrt{2v(\Phi_0)}e^{\phi_0}}{c_0(t)^2M_6^4}\phi_1
\nonumber\\&&\qquad\qquad\qquad=3\left[\frac{\ddot a_0}{a_0}
+\left(\frac{\dot a_0}{a_0}\right)^2\right]+\frac{\ddot c_0}{c_0}
+2\left(\frac{\dot c_0}{c_0}\right)^2+\frac{1}{M_6^4}\tilde p_5\\
\label{CONS5}
&& \tilde p_5'-C_0'\tilde p_6 = 0\\
\label{GAUGE}
&& U'  -\dot Y-\beta e^{\phi_0}e^{C_0(r)}\dot X = 0\\
\label{TRplusRT}
&& \tilde p^t_r = -c_0(t)^2\tilde p^r_t\\
\label{RT}
&& 3\frac{\dot a_0}{a_0}Z-C_0'\dot X 
+\frac{3}{4}(\dot Z-\dot W) =
\frac{c_0(t)^2}{M_6^4}\tilde p^r_t - \frac{\sqrt{2v(\Phi_0)} 
e^{-C_0(r)}}{M_6^4}U -\frac{\dot c_0}{c_0}(W+2\phi_1')\\
\label{CONS1} 
&& \dot{\tilde\rho}+3\frac{\dot a_0}{a_0}(\tilde\rho+\tilde p)
+\frac{\dot c_0}{c_0}(2\tilde\rho+2\tilde p_5 +\tilde p_6)= 
\tilde {p^r_t}'+C_0'\tilde p^r_t \\
\label{PHIEOMPLUSRR}
&&\frac{\phi_1''+C_0'\phi_1'+C_0'W}{3c_0(t)^2}=\left[\frac{\ddot a_0}{a_0}
+\left(\frac{\dot a_0}{a_0}\right)^2-\frac{1}{3}\frac{\ddot c_0}{c_0}
-\frac{\dot a_0}{a_0}\frac{\dot c_0}{c_0}+\frac{\tilde p_5}{3M_6^4}\right]
\eeqa
\beqa
\label{OTHER}
&&\frac{C_0'X'}{c_0(t)^2}-\frac{2X\beta\sqrt{2v(\Phi_0)}
e^{\phi_0}}{M_6^4c_0(t)^2}+\frac{3C_0'W}{2c_0(t)^2}
-\frac{\sqrt{2v(\Phi_0)} e^{-C_0(r)}}{c_0(t)^2M_6^4}Y 
\nonumber\\ && \qquad\qquad = -\frac{1}{4M_6^4}(\tilde\rho-3\tilde p+3\tilde p_6)
+\frac{3}{2}\left[\frac{\ddot a_0}{a_0}
+\left(\frac{\dot a_0}{a_0}\right)^2+\frac{\ddot c_0}{c_0}
+3\frac{\dot a_0}{a_0}\frac{\dot c_0}{c_0}+\frac{4}{3}
\left(\frac{\dot c_0}{c_0}\right)^2\right] 
\eeqa
As was done in \cite{VC}, we assume all the $\rho$'s and $p$'s are 
functions of time only.  Also, in order to simplify calculations and 
because in the absence of a specific microscopic model for the brane 
matter content we are free to choose $\tilde p_6$ as we wish, we will 
assume that 
\beqa
\tilde p_6(r,t) = \theta(r_0-r)e^{2C_0(r)}{\cal P}_6(t)
+ \theta(r-(\pi/k-r_0))e^{2C_0(r)}{\cal P}_{*6}(t).
\eeqa 

\subsection{Solutions to the perturbed equations of motion}
The preceding system of equations may seem daunting, but is actually quite 
straightforward to solve.  One simply uses eqs.\ (\ref{TTminXX}),
(\ref{THminRR}) and(\ref{CONS5}) to solve for $Z(r,t), W(r,t)$ and 
$\tilde p_5(r,t)$ respectively.  Once these solutions are found, one can 
use eq.\ (\ref{PHIEOMPLUSRR}) to solve for $\phi_1(r,t)$ and then either 
eq.\ (\ref{RR}) or eq.\ (\ref{FEOM6}) to solve for $Y(r,t)$.  Then $X(r,t)$ 
can be solved using eq.\ (\ref{OTHER}), while eq.\ (\ref{CONS1}) solves for 
$\tilde p^r_t(r,t)$ and either eq.\ (\ref{GAUGE}) or eq.\ (\ref{RT}) 
solve for $U(r,t)$.  

One must then impose appropriate boundary conditions, namely that all 
solutions are regular at the poles, and smooth across the core/bulk 
boundary except where indicated otherwise by the presence of the 
one-dimensional delta function terms in the stress-energy tensor.  

In the interest of brevity we will only present here the general solutions 
for $W(r,t)$ and $\phi_1(r,t)$ in the bulk, which will be required for 
our argument in the following section.  
\beqa
&& W^{(bulk)}(r,t) = {\cal F}_2^{(bulk)}(t)\sin(kr)\\
&& \phi_1^{(bulk)}(r,t) = \frac{1}{2k}\left[{\cal F}_2^{(bulk)}(t)
+{\cal F}_4^{(bulk)}(t)\right]\ln(1-\cos(kr))
+{\cal F}_5^{(bulk)}(t)+\frac{\cos(kr)}{2k}{\cal F}_2^{(bulk)}(t)
\nonumber\\
&&+ \left[\frac{c_0(t)^2}{k^2}\left(\left(\frac{\ddot c_0}{c_0}\right)
-3\frac{\ddot a_0}{a_0}+3\frac{\dot c_0}{c_0}\frac{\dot a_0}{a_0}
-3\left(\frac{\dot a_0}{a_0}\right)^2\right)\right.
-\left.\frac{1}{2k}\left({\cal F}_2^{(bulk)}(t)+{\cal F}_4^{(bulk)}(t)
\right)
\right]\ln(\sin(kr))\nonumber\\
\eeqa
where ${\cal F}_i(t)$ are constants of integration set by boundary 
conditions.  The interested reader will find solutions to the other 
gauge invariant variables in the appendix.

\section{Presence of the $\lambda_3\neq 0$ solutions}

As we have explained above, the hope is that starting from a static 
$\lambda_3=0$ solution the system will naturally evolve to a static 
$\lambda_3\neq 0$ solution when the brane tensions are perturbed.  In order 
for our results to confirm or rule out this idea, we must verify that our 
perturbative solutions actually include the $\lambda_3\neq 0$ case.  
We will now show this explicitly.

We work in the 
convenient gauge where $B_1(r,t)=0$, which means that effectively $g_{rr} 
= 1$.  In this gauge we can write the metric as 
\beqa
ds^2 = -n(r)dx^{\mu}dx_{\mu}+dr^2
+c(r)^2d\theta^2
\eeqa
and it was shown in \cite{GGP} that
\beqa
\label{l3eqn}
\lambda_3 = \frac{1}{2}c(r)\left[2n(r)^4\phi(r)'
+4n(r)^3n(r)'\right].
\eeqa
In our perturbative language, 
the background solution has a static dilaton and no warping so that 
$\lambda_3$ vanishes identically at this order.  If we wish to prove 
that our perturbative results include the nonconical solutions with 
$\lambda_3\neq 0$, we have to show that at ${\cal O}(\rho)$, the combination 
on the right hand side of eq.\ (\ref{l3eqn}) does not vanish.  For static 
perturbations, we can write eq.\ (\ref{l3eqn}) to ${\cal O}(\rho)$ as
\beqa
\delta\lambda_3 = \frac{1}{2}e^{C_0(r)}\left[2\phi_1(r)'+W(r)\right].
\eeqa
Substituting our bulk solutions into this expression and assuming staticity 
(i.e. $a_0(t)\equiv 1, c_0(t)\equiv 1, \rho = -p$ and 
$\rho_* = -p_*$), we find that 
\beqa
\delta\lambda_3 =\frac{1}{12k^4M_6^4}\left(1-\cos(kr_0)\right)^3
\left({\cal P}_6(t)-{\cal P}_{*6}(t)\right).
\eeqa
This result tells us that the solutions with nonconical singularities 
at the poles are related to the presence on the branes of stress-energy 
along the extra dimensions (recall that we defined 
$s^{\theta}_{\theta}-s^r_r = \theta(r_0-r) p_6(r,t) \equiv 
\theta(r_0-r) e^{2C_0(r)}{\cal P}_6(t)$ and 
${s_*}^{\theta}_{\theta}-{s_*}^r_r =  \theta(r-r_*) p_{*6}(r,t) 
\equiv \theta(r-r_*)e^{2C_0(r)}{\cal P}_{*6}(t)$ ).  This result is 
compatible with the discussions of \cite{Burgess:2004dh} and 
\cite{navarrosantiago}.  

Having shown that our solutions include the case of interest, 
we can now address the question of whether the model 
can solve the cosmological constant problem through self-tuning.

\section{Cosmology and the question of self-tuning}

\subsection{Boundary conditions and the Friedmann equations}

In braneworld models, one derives the Friedmann equations on the brane(s) 
by imposing appropriate boundary conditions.  In the model we are studying, 
these imply that all functions are regular at the origin, and smooth across 
the brane/bulk boundaries.  Since all the gauge invariant quantities we 
have defined only appear as first $r$ derivatives in the equations of 
motion, the functions themselves must match, but not their first 
derivatives.  The only exception is $\phi_1(r,t)$, for which both the 
function itself and its first derivative must be matched across the 
boundaries at $r=r_0$ and $r=\pi/k-r_0$.  One must also take into account 
the one-dimensional delta functions in the brane stress-energy 
tensor, which affect the following junction conditions
\beqa
\lim_{\epsilon\rightarrow 0}X(r_0-\epsilon,t)-X(r_0+\epsilon,t) &=& 
\frac{\tan(kr_0)}{kM_6^4}c_0(t)^2{\cal F}_0(t)\\
\lim_{\epsilon\rightarrow 0}X(r_*+\epsilon,t)-X(r_*-\epsilon,t) &=& 
\frac{\tan(kr_0)}{kM_6^4}c_0(t)^2{\cal F}_*(t)\\
p^r_t(r_0,t) &=& -\dot{\cal F}_0(t) - 2\frac{\dot c_0}{c_0}{\cal F}_0(t)\\
{p^r_t}_*(r_*,t) &=& \dot{\cal F}_*(t) + 2\frac{\dot c_0}{c_0}{\cal F}_*(t)
\eeqa
Solving for all variables and boundary conditions, one is led to the 
Friedmann equations, and conservation of energy
\beqa
\left(\frac{\dot a_0}{a_0}\right)^2 &=& \frac{(1-\cos(kr_0))}{8M_6^4}
(\rho+p+\rho_*+p_*)-
\frac{(1-\cos(kr_0))^3}{72k^2M_6^4}({\cal P}_6+{{\cal P}_*}_6)\nonumber\\
&+&\frac{2}{3}\frac{\ddot c_0}{c_0}+\left(\frac{\dot c_0}{c_0}\right)^2\\
\frac{\ddot a_0}{a_0}-\left(\frac{\dot a_0}{a_0}\right)^2&=&
-\frac{(1-\cos(kr_0))}{4M_6^4}(\rho+p+\rho_*+p_*)
-\frac{\ddot c_0}{c_0}-2\left(\frac{\dot c_0}{c_0}\right)^2
+\frac{\dot a_0}{a_0}\frac{\dot c_0}{c_0}\\
\dot\rho +3\frac{\dot a_0}{a_0}(\rho+p)&=&
-\frac{k\sin(kr_0)}{1-\cos(kr_0)}\left(\dot {\cal F}_0
+2\frac{\dot c_0}{c_0}{\cal F}_0\right)-\frac{\dot c_0}{c_0}\left(2\rho 
+\frac{(1-\cos(kr_0))^2}{3k^2}{\cal P}_6\right)\\
\dot\rho_* +3\frac{\dot a_0}{a_0}(\rho_*+p_*)&=&
-\frac{k\sin(kr_0)}{1-\cos(kr_0)}\left(\dot {\cal F}_*
+2\frac{\dot c_0}{c_0}{\cal F}_*\right)-\frac{\dot c_0}{c_0}\left(2\rho_* 
+\frac{(1-\cos(kr_0))^2}{3k^2}{{\cal P}_*}_6\right)
\eeqa
Consistency between these four equations implies that
\beqa
\label{consist}
{\cal F}_0+{\cal F}_* &=& -\frac{(1-\cos(kr_0))}{4k\sin(kr_0)}
(\rho-3p+\rho_*-3p_*)\nonumber\\
&+&\frac{15\cos(kr_0)-6\cos(2kr_0)-10+\cos(3kr_0)}{48k^3\sin(kr_0)}
({\cal P}_6+{{\cal P}_*}_6)+\frac{{\cal Q}_1}{c_0(t)^2}\nonumber\\&+&
\frac{4M_6^4}{k\sin(kr_0)}\left[
\frac{\ddot c_0}{c_0}+\left(\frac{\dot c_0}{c_0}\right)^2
+3\frac{\dot a_0}{a_0}\frac{\dot c_0}{c_0}\right].
\eeqa
Note that the constant 
of integration ${\cal Q}_1$ which appears in eq.\ (\ref{consist}) can be 
shown to be proportional to the perturbation to the flux.


\subsection{Effective four dimensional quantities}

As was discussed in \cite{VC}, the quantities we have used so far are not 
the ones a four dimensional observer on the brane would identify as the 
energy density and pressure.  Indeed, to define the effective four 
dimensional quantities, we must integrate the 6D quantities over 
the thickness of the brane,
\beqa
{S^{(4)}}^a_b = 2\pi\int_0^{r_0} dr\, b(r,t)\,c(r,t)\, {S^{(6)}}^a_b
\eeqa
which with our choice of gauge ($g_{rr}\equiv 1$) leads to
\beqa
\rho^{(4)}(t) &=& 2\pi\int_{0}^{r_0}c_0(t)^2 e^{C_0(r)}\rho(t) dr  
+2\pi c_0(t)^2 e^{C_0(r_0)}{\cal F}_0(t)\\
p^{(4)}(t) &=& 2\pi\int_{0}^{r_0}c_0(t)^2 e^{C_0(r)}p(t) dr 
-2\pi c_0(t)^2 e^{C_0(r_0)}{\cal F}_0(t)\\
\rho_*^{(4)}(t) &=& 2\pi\int_{\pi/k-r_0}^{ \pi/k}c_0(t)^2  
e^{C_0(r)}\rho_*(t) dr +2\pi c_0(t)^2 e^{C_0(r_*)}{\cal F}_*(t)\\
p_*^{(4)}(t) &=& 2\pi\int_{\pi/k-r_0}^{ \pi/k } c_0(t)^2 
e^{C_0(r)}p_*(t) dr -2\pi c_0(t)^2 e^{C_0(r_*)}{\cal F}_*(t).
\eeqa
We also define the 4d Newton constant as
\beqa
\frac{1}{8\pi G_4(t)} &=& M_4^2 = 2\pi\int^{\pi/k}_{0} c_0(t)^2 e^{C_0(r)} 
M_6^4dr\\
 \Rightarrow G_4(t)&=& \frac{k^2}{32\pi^2M_6^4c_0(t)^2}.
\eeqa

We come now to the main result of this paper, the Friedmann equations 
on general codimension-two branes in six-dimensional supergravity.  
Writing these in terms of the effective four dimensional quantities 
defined above, we find
\beqa
\label{fried1}
\left(\frac{\dot a_0}{a_0}\right)^2&=&\frac{8\pi G_4(t)}{3}
(\rho^{(4)}+\rho_*^{(4)})-\frac{16\pi^2\sin(kr_0)}{3k}G_4(t){\cal Q}_1+
\frac{1}{3}\left(\frac{\dot c_0}{c_0}\right)^2\nonumber\\
&-&2\frac{\dot a_0}{a_0}\frac{\dot c_0}{c_0}\\
\label{fried2}
\frac{\ddot a_0}{a_0}-\left(\frac{\dot a_0}{a_0}\right)^2&=&-4\pi G_4(t)
(\rho^{(4)}+p^{(4)}+\rho^{(4)}_*+p^{(4)}_*)
-\frac{\ddot c_0}{c_0}-2\left(\frac{\dot c_0}{c_0}\right)^2\nonumber\\
&+&\frac{\dot a_0}{a_0}\frac{\dot c_0}{c_0}\\
\label{cons1}
\dot\rho^{(4)} &=& -3\frac{\dot a_0}{a_0}(\rho^{(4)}+p^{(4)})
-\frac{2\pi(1-\cos(kr_0))^3c_0(t)\dot c_0(t)}{3k^4}{\cal P}_6\\
\label{cons2}
\dot\rho_*^{(4)} &=& -3\frac{\dot a_0}{a_0}(\rho_*^{(4)}+p_*^{(4)})
-\frac{2\pi(1-\cos(kr_0))^3c_0(t)\dot c_0(t)}{3k^4}{{\cal P}_*}_6\\
\label{fried3}
\frac{\ddot c_0}{c_0}+\left(\frac{\dot c_0}{c_0}\right)^2
+3\frac{\dot a_0}{a_0}\frac{\dot c_0}{c_0}&=&
\pi G_4(t)\left(\rho^{(4)}-3p^{(4)}
+\rho_*^{(4)}-3p_*^{(4)}\right)\nonumber\\
&+&\frac{2\pi^2c_0(t)^2(1-\cos(kr_0))^3}{3k^4}G_4(t)\left({\cal P}_6
+{{\cal P}_*}_6\right)-\frac{8\pi^2\sin(kr_0)}{k}G_4(t){\cal Q}_1.\nonumber\\
\eeqa

We see a major difference with the Einstein-Hilbert case studied in 
\cite{VC}: here the 4d Newton constant is time-dependent at first order 
in the perturbations, whereas in the Einstein-Hilbert case such corrections 
would only appear at ${\cal O}(\rho^2)$.  Although we will 
not focus on this point for the rest of the paper, we could use results of 
fifth force experiments and measures of the time variation of Newton's 
constant to strongly constrain the dynamics of the variable $c_0(t)$.  
In fact, one can show from the equations presented above that in the 
${\cal P}_6 = {\cal P}_{*6} = 0$ limit, the equations can be derived from the 
effective action 
\beqa
S = \int d^4x \sqrt{-g}\left(\frac{1}{16\pi \bar G}\left[\zeta{\cal R}-\frac{1}{2\zeta}\partial_{\mu}\zeta\partial^{\mu}\zeta-V(\zeta)\right]+{\cal L}_m\right)
\eeqa
by making the identifications $c_0(t)=\sqrt{\zeta}$, $V(\zeta)=-32\pi^2\sin(kr_0)\bar G {\cal Q}_1/k$ and $\bar G/\zeta(t) = G_4(t)$
In other words, the model in that case corresponds to 
Brans-Dicke gravity with $\omega = 1/2$ and a flat potential, a theory 
which is clearly ruled out by experiment.  Note that this is consistent 
with the arguments of \cite{GP}, since this is precisely the limit 
where non-conical singularities are absent, and corresponds to the case 
that was studied in that work.  

When one does include non-zero ${\cal P}_6$ and ${\cal P}_{*6}$ terms, we 
see from eqs.(\ref{cons1},\ref{cons2}) that conservation of energy is 
violated from a four dimensional perspective, and it is thus hard to 
imagine how one might write down an effective four dimensional action.  It 
should be noted that this should not come as a surprise, since it is precisely 
this type of intrinsically 
extra dimensional effect which motivated this approach as a plausible 
way to circumvent Weinberg's no-go theorem and arrive at a self-tuning 
solution to the cosmological constant problem.  

\subsection{Absence of self-tuning}

Using eqs.\ (\ref{fried1})-(\ref{fried3}), we can now categorically answer in 
the negative the question of whether or not the model self-tunes to a 
static solution under perturbations to the brane tensions.  

Indeed, it was 
already known that in the absence of the solutions with nonconical 
singularities, one needs to fine-tune the brane tensions with the flux 
from the bulk gauge field in order to obtain static solutions 
\cite{Aghababaie:2003wz,Aghababaie:2003ar,GGP,GP,Burgess:2004dh}.  
The hope, as explained above, was that including the solutions with 
nonconical singularities would allow the model to naturally evolve from 
a static solution with conical singularities to a static solution with 
nonconical singularities when one perturbs the tensions.  We see clearly 
now that this cannot happen, for two reasons.  

First, nonconical singularities are only present when the functions 
${\cal P}_6$ and ${\cal P}_{*6}$ are nonzero.  However, these have no 
dynamical equation of motion; they are free parameters in this model.  
Therefore, even if it were possible to use them in order to cancel a change 
in the brane tensions, this would have to be done by hand and would thus 
represent an arbitrary fine-tuning of the model's parameters.  This would 
be made all the more unnatural by the fact that it would have to be a 
time-dependent fine-tuning if it were to hold across multiple phase 
transitions on either brane.  

Second, it is not even possible to use the nonconical singularities to cancel perturbations to the brane tensions and get new static 
solutions.  This can be seen by setting $\rho^{(4)} = -p^{(4)}$ and 
$\rho_*^{(4)} = -p_*^{(4)}$ in the above equations, and setting all 
time derivatives to zero.  We then find that eq.\ (\ref{fried1}) necessitates 
a relation between $\rho^{(4)}$, $\rho^{(4)}_*$ and ${\cal Q}_1$ that, 
when plugged into eq.\ (\ref{fried3}) tells us that 
${\cal P}_6 = -{\cal P}_{*6}$.  In other words, one must tune the brane 
tensions in order to obtain a static solution regardless of whether we have 
conical or nonconical singularities.  Moreover, if we do have nonconical 
singularities, we must perform an {\it additional} fine-tuning of the 
functions ${\cal P}_6$ and ${\cal P}_{*6}$ sourcing them in order to keep the solutions static.  Therefore, 
we have conclusively shown that no self-tuning mechanism for the 
cosmological constant is present in the type of six dimensional 
supergravity we have considered here.  

We should note that although this final result agrees with the conclusions 
of \cite{GP}, the analysis we have performed here was far from superfluous, 
since we have shown that self-tuning fails not only for the simple case 
considered in that work, but in the much more general case which includes 
warping and nonconical singularities.  Indeed, the solutions we find are 
qualitatively different from the ones in \cite{GP} when ${\cal P}_6$ and 
${\cal P}_{*6}$ are nonzero.  One cannot simply derive them from an 
effective four dimensional action since they correspond to a theory where 
the ususal law of conservation of energy does not hold (see 
eqs.\ (\ref{cons1},\ref{cons2})).

One final point of interest is the fact that one can see from our results 
that we cannot set only one of $a_0$ and $c_0$ constant without imposing 
unnatural constraints on the brane stress-energy tensor components (see 
eq.(\ref{fried3}) in the case $c_0=cst$ for example).  They must either 
both be constant (which requires the fine-tuning mentioned in the previous 
paragraphs), or both be time dependent.  This explains why previous 
solutions always singled out solutions with Minkowski 4-space over 
de Sitter or anti-de Sitter space: the initial assumption was always made that the 
internal space was static!  Indeed, rather than saying that once the radion  
and dilaton are stabilized, solutions with a static external space are 
required by the equations of motion, a more correct statement would be that 
choosing the large dimensions to be Minkowski is actually part of the tuning 
one has to do in order to obtain a static radion and dilaton.

\subsection{Dropping axial symmetry: a loophole?}

Although our results appear quite conclusive regarding the absence of a 
self-tuning mechanism for the cosmological constant in six dimensional 
supergravity, one should be keep in mind which assumptions were
made in deriving this result, to see if relaxing one of them 
might in fact lead to a different conclusion.  

The solutions we have found are more general than those that had so far 
been presented in the literature because they allow for general equations 
of state for the brane content and they include the possibility of warping 
and of nonconical singularities in a fully dynamical context.  There is 
however one restriction we have imposed on our metric ansatz which could be 
relaxed: axial symmetry of the internal space.  

In a recent paper 
\cite{Redi}, explicit solutions have been found with an arbitrary number of 
3-branes 
on a compact two-dimensional internal space by dropping the requirement 
of axial symmetry.  The analysis in \cite{Redi} concludes that the only 
solution with two branes is one with equal tension branes located at the 
poles of a spherical internal space, but this result is derived by 
excluding the possibility of warping and is 
restricted to Einstein-Hilbert gravity.  It is interesting to ask 
what might happen in the supersymmetric case if we were to add the possibility 
of breaking axial symmetry in response to a change in the brane tensions, in 
addition to the possibility of nonconical singularities and warping.  
While there is no compelling reason to believe that allowing the branes to 
shift their positions would lead to 
self-tuning, we feel it is nonetheless an avenue worth investigating in 
future work, since it could conceivably lead to interesting and 
perhaps unexpected effects.

\section{Conclusion and outlook}

In this paper, we have studied the cosmology on a codimension-two brane 
in six-dimensional supergravity.  We achieved this by solving the dynamical 
field equations linearized around a static background.  In order to deal 
with the mildly singular behaviour of the warp factor at the position of the 
branes when the equation of state of their matter content is different from 
that of pure tension, we regularized the branes by giving them a nonzero 
thickness.  

Our results show that there is no self-tuning of the cosmological constant 
to zero in such a setup.  Rather, any change to the branes' tensions leads 
to expansion in both the external space and the internal space.  In fact, 
we find that the internal and external space must always either both be static 
(which necessitates fine-tuning) or both be time dependent, but that we cannot 
have one be static while the 
other evolves without imposing unnatural relations between the brane stress-energy 
tensor components.  This explains why previous solutions, which always assumed a 
static internal space from the start, singled out static solutions in the large 
dimensions.  

Our solutions include both nonconical singularities and warping of the metric, 
and we emphasize that this represents a new result, since no dynamical 
treatment of such a model including both these possibilities had been carried out 
to date.  The only assumption we made is that of axial symmetry of the internal space, 
and although it would be interesting to look for solutions where this assumption is 
dropped, we see no reason to believe that it would lead to self-tuning.  

Although we do not believe that 6D supergravity theories offer a solution to
the cosmological constant problem, such models with large extra dimensions can
nevertheless provide a novel source of dynamical dark energy (whose size is
naturally of the right magnitude once the cosmological constant is set to zero)
as well as accelerator signals which would be visible at the LHC
\cite{Burgess:2004ib}, and they therefore merit more study along these lines.

\section{Acknowledgments}
We would like to thank Cliff Burgess for many discussions on this topic.  
We are supported in part by Canada's National Sciences and Engineering 
Research Council and FQRNT of Qu\'ebec.

\appendix
\section{General solutions to the perturbed equations of motion}
We present here the general solutions to the perturbed equations of motion 
(\ref{THminRR})-(\ref{OTHER}).  We only present the solutions in the 
core located in the region $0\leq r\leq r_0$.  Bulk solutions have the same 
form, without the terms coming from the brane stress-energy, while solutions 
in the other core have the same form with the terms from the brane 
stress-energy replaced by the corresponding terms for this second brane 
($\rho_*$ instead of $\rho$ for example).  The solutions for the 
integration ``constants'' ${\cal F}_i(t)$ will not be given explicitly, 
but can be found in each region of the model through the imposition of 
boundary conditions, as explained in the main text.

\beqa
Z(r,t) &=& \frac{\cos(kr)c_0(t)^2}{k\sin(kr)}\left[
2\left(\frac{\dot a_0}{a_0}\right)^2-2\frac{\ddot a_0}{a_0}
-2\frac{\ddot c_0}{c_0}+2\frac{\dot a_0}{a_0}\frac{\dot c_0}{c_0}
-4\left(\frac{\dot c_0}{c_0}\right)^2-\frac{1}{M_6^4}(\rho(t)+p(t))\right]
\nonumber\\&+&\frac{{\cal F}_1(t)}{\sin(kr)}\\
W(r,t) &=& -\frac{\cos(kr)\sin(kr)c_0(t)^2}{k^3M_6^4}{\cal P}_6(t) 
+ \sin(kr){\cal F}_2(t)\\
p^r_r(r,t) &=& -\frac{\cos(2kr)}{4k^2}{\cal P}_6(t)+{\cal F}_3(t)\\
\phi_1(r,t) &=& \ln(\sin(kr))\frac{c_0(t)^2}{k^2}\left[
-3\left(\frac{\dot a_0}{a_0}\right)^2
+3\frac{\dot a_0}{a_0}\frac{\dot c_0}{c_0}
-3\frac{\ddot a_0}{a_0}+\frac{\ddot c_0}{c_0}
-\frac{k{\cal F}_2(t)}{2c_0(t)^2}-\frac{5{\cal P}_6(t)}{12k^2M_6^4}\right.
\nonumber\\&-&\left.\frac{k{\cal F}_4(t)}{c_0(t)^2}
-\frac{{\cal F}_3(t)}{M_6^4}\right]
+\frac{{\cal F}_2(t)}{2k}\cos(kr)
-\frac{c_0(t)^2{\cal P}_6(t)}{12k^4M_6^4}\cos(kr)^2+{\cal F}_5(t)\nonumber\\
&+&\frac{1}{k}\ln(1-\cos(kr))\left[\frac{{\cal F}_2(t)}{2}
+{\cal F}_4(t)\right]\\
Y(r,t) &=& \sin(kr)\ln(\sin(kr))\frac{\beta e^{\phi_0}c_0(t)^2}{12k^3}\left[
36\frac{\ddot a_0}{a_0}+36\left(\frac{\dot a_0}{a_0}\right)^2
-36\frac{\dot a_0}{a_0}\frac{\dot c_0}{c_0}-12\frac{\ddot c_0}{c_0}
+\frac{12{\cal F}_3(t)}{M_6^4}\right.\nonumber\\
&+&\left.5\frac{{\cal P}_6(t)}{k^2M_6^4}+12\frac{k{\cal F}_4(t)}{c_0(t)^2}
+6\frac{k{\cal F}_2(t)}{c_0(t)^2}\right]
+\frac{7\beta e^{\phi_0}c_0(t)^2{\cal P}_6(t)}{12k^5M_6^4}\sin(kr)\cos(kr)^2
\nonumber\\&-&\frac{3\beta e^{\phi_0}{\cal F}_2(t)}{2k^2}\sin(kr)\cos(kr)
-\frac{\beta e^{\phi_0}({\cal F}_2(t)+2{\cal F}_4(t))}
{2k^2}\sin(kr)\ln(1-\cos(kr))\nonumber\\
&+&\sin(kr)\frac{\beta e^{\phi_0}c_0(t)^2}{4k^3}\left[12\frac{\ddot a_0}{a_0}
+12\left(\frac{\dot a_0}{a_0}\right)^2
+12\frac{\dot a_0}{a_0}\frac{\dot c_0}{c_0}+4\frac{\ddot c_0}{c_0}
+8\left(\frac{\dot c_0}{c_0}\right)^2
+4\frac{{\cal F}_3(t)}{M_6^4}\right.\nonumber\\
&+&\left.\frac{{\cal P}_6(t)}{k^2M_6^4}-4\frac{k^2{\cal F}_5(t)}{c_0(t)^2}\right]\\
X(r,t) &=& \tan(kr)^2\ln(\sin(kr))\frac{c_0(t)^2}{24k^2}\left[
36\frac{\ddot a_0}{a_0}+36\left(\frac{\dot a_0}{a_0}\right)^2
-36\frac{\dot a_0}{a_0}\frac{\dot c_0}{c_0}-12\frac{\ddot c_0}{c_0}
+\frac{12{\cal F}_3(t)}{M_6^4}\right.\nonumber\\
&+&\left.5\frac{{\cal P}_6(t)}{k^2M_6^4}+12\frac{k{\cal F}_4(t)}{c_0(t)^2}
+6\frac{k{\cal F}_2(t)}{c_0(t)^2}\right]-\frac{2{\cal F}_4(t)
+{\cal F}_2(t)}{4k}\tan(kr)^2\ln(1-\cos(kr))\nonumber\\
&-&\frac{17c_0(t)^2{\cal P}_6(t)}{24M_6^4k^4}\cos(kr)^2
+\frac{{\cal F}_2(t)}{k}\cos(kr)-\frac{2{\cal F}_4(t)+{\cal F}_2(t)}{4k\cos(kr)}\nonumber\\
&+&\frac{3c_0(t)^2}{4k^2\cos(kr)^2}\left[-\frac{\ddot a_0}{a_0}
-\left(\frac{\dot a_0}{a_0}\right)^2
+\frac{\dot a_0}{a_0}\frac{\dot c_0}{c_0}+\frac{1}{3}\frac{\ddot c_0}{c_0}
+\frac{4k^2{\cal F}_6(t)}{3c_0(t)^2}+\frac{k{\cal F}_2(t)}{c_0(t)^2}\right.\nonumber\\
&+&\left.\frac{2k{\cal F}_4(t)}{3c_0(t)^2}
-\frac{{\cal F}_3(t)}{3M_6^4}-\frac{7{\cal P}_6(t)}{18k^2M_6^4}\right]
+\frac{3c_0(t)^2}{2k^2}\left[-\frac{\ddot a_0}{a_0}-\left(\frac{\dot a_0}{a_0}\right)^2
-3\frac{\dot a_0}{a_0}\frac{\dot c_0}{c_0}-\frac{\ddot c_0}{c_0}\right.\nonumber\\
&-&\left.\frac{4}{3}\left(\frac{\dot c_0}{c_0}\right)^2-\frac{{\cal F}_39t)}{6M_6^4}+\frac{k^2{\cal F}_5(t)}{3c_0(t)^2}
+\frac{17{\cal P}_6(t)}{72k^2M_6^4}+\frac{1}{M_6^4}(\rho(t)-3p(t))\right]\\
p^r_t(r,t) &=& \frac{{\cal P}_6(t)}{3k^3}\frac{\dot c_0}{c_0}\cot(kr)(2\cos(kr)^2-1)
-\frac{\cot(kr)}{k}\left[\dot\rho+3\frac{\dot a_0}{a_0}(\rho(t)+p(t))\right.\nonumber\\
&+&\left.\frac{\dot c_0}{c_0}\left(2\rho(t)+2{\cal F}_3(t)+\frac{7{\cal P}_6(t)}{6k^2}\right)\right]
+\frac{{\cal F}_7(t)}{\sin(kr)}
\eeqa
We have not written down $U(r,t)$ explicitly since it can be obtained 
trivially from the algebraic equation (\ref{RT}) once all other functions 
are given.


\begin{thebibliography}{999}

\bibitem{Arkani-Hamed:2000eg}
N.~Arkani-Hamed, S.~Dimopoulos, N.~Kaloper and R.~Sundrum,
``A small cosmological constant from a large extra dimension,''
Phys.\ Lett.\ B {\bf 480}, 193 (2000)
[arXiv:hep-th/0001197].

\bibitem{Kachru:2000hf}
S.~Kachru, M.~B.~Schulz and E.~Silverstein,
``Self-tuning flat domain walls in 5d gravity and string theory,''
Phys.\ Rev.\ D {\bf 62}, 045021 (2000)
[arXiv:hep-th/0001206].

\bibitem{Forste:2000ft}
S.~Forste, Z.~Lalak, S.~Lavignac and H.~P.~Nilles,
``The cosmological constant problem from a brane-world perspective,''
JHEP {\bf 0009}, 034 (2000)
[arXiv:hep-th/0006139].

\bibitem{Chen:2000at}
J.~W.~Chen, M.~A.~Luty and E.~Ponton,
``A critical cosmological constant from millimeter extra dimensions,''
JHEP {\bf 0009}, 012 (2000)
[arXiv:hep-th/0003067].

\bibitem{Carroll}
S.~M.~Carroll and M.~M.~Guica,
``Sidestepping the cosmological constant with football-shaped extra  dimensions,''
arXiv:hep-th/0302067.

\bibitem{Navarro1}
I.~Navarro,
``Codimension two compactifications and the cosmological constant  problem,''
JCAP {\bf 0309}, 004 (2003)
[arXiv:hep-th/0302129].

\bibitem{Gunther:2003zn}
U.~Gunther, P.~Moniz and A.~Zhuk,
``Nonlinear multidimensional cosmological models with form fields:
Stabilization of extra dimensions and the cosmological constant  problem,''
Phys.\ Rev.\ D {\bf 68}, 044010 (2003)
[arXiv:hep-th/0303023].

\bibitem{Cline:2003ak}
J.~M.~Cline, J.~Descheneau, M.~Giovannini and J.~Vinet,
``Cosmology of codimension-two braneworlds,''
JHEP {\bf 0306}, 048 (2003)
[arXiv:hep-th/0304147].

\bibitem{Aghababaie:2003wz}
Y.~Aghababaie, C.~P.~Burgess, S.~L.~Parameswaran and F.~Quevedo,
``Towards a naturally small cosmological constant from branes in 6D
supergravity,''
Nucl.\ Phys.\ B {\bf 680}, 389 (2004)
[arXiv:hep-th/0304256].

\bibitem{Navarro2}
I.~Navarro,
``Spheres, deficit angles and the cosmological constant,''
Class.\ Quant.\ Grav.\  {\bf 20}, 3603 (2003)
[arXiv:hep-th/0305014].

\bibitem{Kehagias}
A.~Kehagias,
``A conical tear drop as a vacuum-energy drain for the solution of the
cosmological constant problem,''
Phys.\ Lett.\ B {\bf 600}, 133 (2004)
[arXiv:hep-th/0406025].

\bibitem{VC}
J.~Vinet and J.~M.~Cline,
``Can codimension-two branes solve the cosmological constant problem?,''
Phys.\ Rev.\ D {\bf 70}, 083514 (2004)
[arXiv:hep-th/0406141].

\bibitem{Weinberg}
S.~Weinberg,
``The Cosmological Constant Problem,''
Rev.\ Mod.\ Phys.\  {\bf 61}, 1 (1989).

\bibitem{Binetruy:2000wn}
P.~Binetruy, J.~M.~Cline and C.~Grojean,
``Dynamical instability of brane solutions with a self-tuning  cosmological
constant,''
Phys.\ Lett.\ B {\bf 489}, 403 (2000)
[arXiv:hep-th/0007029].

\bibitem{Csaki:2000dm}
C.~Csaki, J.~Erlich and C.~Grojean,
``Gravitational Lorentz violations and adjustment of the cosmological
constant in asymmetrically warped spacetimes,''
Nucl.\ Phys.\ B {\bf 604}, 312 (2001)
[arXiv:hep-th/0012143].

\bibitem{Csaki:2001mn}
C.~Csaki, J.~Erlich and C.~Grojean,
``Essay on gravitation: The cosmological constant problem in brane-worlds  and
gravitational Lorentz violations,''
Gen.\ Rel.\ Grav.\  {\bf 33}, 1921 (2001)
[arXiv:gr-qc/0105114].

\bibitem{Cline:2001yt}
J.~M.~Cline and H.~Firouzjahi,
``No-go theorem for horizon-shielded self-tuning singularities,''
Phys.\ Rev.\ D {\bf 65}, 043501 (2002)
[arXiv:hep-th/0107198].

\bibitem{Sundrum:1998ns}
R.~Sundrum,
``Compactification for a three-brane universe,''
Phys.\ Rev.\ D {\bf 59}, 085010 (1999)
[arXiv:hep-ph/9807348].

\bibitem{GS}
T.~Gherghetta and M.~Shaposhnikov,
``Localizing gravity on a string-like defect in six dimensions,''
Phys.\ Rev.\ Lett.\  {\bf 85}, 240 (2000)
[arXiv:hep-th/0004014].

\bibitem{Nihei:2000gb}
T.~Nihei,
``Gravity localization with a domain wall junction in six dimensions,''
Phys.\ Rev.\ D {\bf 62}, 124017 (2000)
[arXiv:hep-th/0005014].

\bibitem{Ponton:2000gi}
E.~Ponton and E.~Poppitz,
``Gravity localization on string-like defects in codimension two and the  AdS/CFT correspondence,''
JHEP {\bf 0102}, 042 (2001)
[arXiv:hep-th/0012033].

\bibitem{Giovannini:2001hh} 
M.~Giovannini, H.~Meyer and M.~E.~Shaposhnikov, 
``Warped compactification on Abelian vortex in six dimensions,'' 
Nucl.\ Phys.\ B {\bf 619}, 615 (2001) [arXiv:hep-th/0104118]. 

\bibitem{Kanti:2001vb}
P.~Kanti, R.~Madden and K.~A.~Olive,
``A 6-D brane world model,''
Phys.\ Rev.\ D {\bf 64}, 044021 (2001)
[arXiv:hep-th/0104177].

\bibitem{Neupane:2001kd}
I.~P.~Neupane,
``Localized gravity with higher curvature terms,''
Class.\ Quant.\ Grav.\  {\bf 19}, 5507 (2002)
[arXiv:hep-th/0106100].

\bibitem{chacko}
Z.~Chacko, P.~J.~Fox, A.~E.~Nelson and N.~Weiner,
``Large Extra Dimensions from a Small Extra Dimension,"
JHEP {\bf 0203} (2002) 001
[arXiv:hep-ph/0106343].

\bibitem{KMPR}
I.~I.~Kogan, S.~Mouslopoulos, A.~Papazoglou and G.~G.~Ross,
``Multigravity in six dimensions: Generating bounces with flat positive
tension branes,''
Phys.\ Rev.\ D {\bf 64}, 124014 (2001)
[arXiv:hep-th/0107086].

\bibitem{Carroll:2001ih}
S.~M.~Carroll, J.~Geddes, M.~B.~Hoffman and R.~M.~Wald,
``Classical stabilization of homogeneous extra dimensions,''
Phys.\ Rev.\ D {\bf 66}, 024036 (2002)
[arXiv:hep-th/0110149].

\bibitem{BCCF}
C.~P.~Burgess, J.~M.~Cline, N.~R.~Constable and H.~Firouzjahi,
JHEP {\bf 0201} (2002) 014
[arXiv:hep-th/0112047].

\bibitem{Corradini:2002ta}
O.~Corradini, A.~Iglesias, Z.~Kakushadze and P.~Langfelder,
``A remark on smoothing out higher codimension branes,''
Mod.\ Phys.\ Lett.\ A {\bf 17}, 795 (2002)
[arXiv:hep-th/0201201].

\bibitem{GGP}
G.~W.~Gibbons, R.~Guven and C.~N.~Pope,
``3-branes and uniqueness of the Salam-Sezgin vacuum,''
Phys.\ Lett.\ B {\bf 595}, 498 (2004)
[arXiv:hep-th/0307238].

\bibitem{Aghababaie:2003ar}
Y.~Aghababaie {\it et al.},
``Warped brane worlds in six dimensional supergravity,''
JHEP {\bf 0309}, 037 (2003)
[arXiv:hep-th/0308064].

\bibitem{Nilles:2003km}
H.~P.~Nilles, A.~Papazoglou and G.~Tasinato,
``Selftuning and its footprints,''
Nucl.\ Phys.\ B {\bf 677}, 405 (2004)
[arXiv:hep-th/0309042].

\bibitem{Lee:2003wg}
H.~M.~Lee,
``A comment on the self-tuning of cosmological constant with deficit  angle on
a sphere,''
Phys.\ Lett.\ B {\bf 587}, 117 (2004)
[arXiv:hep-th/0309050].

\bibitem{deCarlos:2003nq}
B.~de Carlos and J.~M.~Moreno,
``A cigar-like universe,''
JHEP {\bf 0311}, 040 (2003)
[arXiv:hep-th/0309259].

\bibitem{Gregory2}
P.~Bostock, R.~Gregory, I.~Navarro and J.~Santiago,
``Einstein gravity on the codimension 2 brane?,''
Phys.\ Rev.\ Lett.\  {\bf 92}, 221601 (2004)
[arXiv:hep-th/0311074].

\bibitem{Burgess:2004kd}
C.~P.~Burgess,
``Supersymmetric large extra dimensions and the cosmological constant: An
update,''
Annals Phys.\  {\bf 313}, 283 (2004)
[arXiv:hep-th/0402200].

\bibitem{Graesser:2004xv}
M.~L.~Graesser, J.~E.~Kile and P.~Wang,
``Gravitational perturbations of a six dimensional self-tuning model,''
Phys.\ Rev.\ D {\bf 70}, 024008 (2004)
[arXiv:hep-th/0403074].

\bibitem{Burgess:2004yq}
C.~P.~Burgess, J.~Matias and F.~Quevedo,
``MSLED: A minimal supersymmetric large extra dimensions scenario,''
arXiv:hep-ph/0404135.

\bibitem{kannosoda}
S.~Kanno and J.~Soda,
``Quasi-thick codimension 2 braneworld,''
JCAP {\bf 0407}, 002 (2004)
[arXiv:hep-th/0404207].

\bibitem{Wu:2004gp}
Z.~C.~Wu,
``No-boundary codimension-two braneworld,''
arXiv:hep-th/0405249.

\bibitem{GP}
J.~Garriga and M.~Porrati,
``Football shaped extra dimensions and the absence of self-tuning,''
JHEP {\bf 0408}, 028 (2004)
[arXiv:hep-th/0406158].

\bibitem{Burgess:2004xk}
C.~P.~Burgess,
``Supersymmetric large extra dimensions,''
arXiv:hep-ph/0406214.

\bibitem{Burgess:2004dh}
C.~P.~Burgess, F.~Quevedo, G.~Tasinato and I.~Zavala,
``General axisymmetric solutions and self-tuning in 6D chiral gauged
supergravity,''
JHEP {\bf 0411}, 069 (2004)
[arXiv:hep-th/0408109].

\bibitem{Burgess:2004ib}
C.~P.~Burgess,
``Towards a natural theory of dark energy: Supersymmetric large extra
dimensions,''
arXiv:hep-th/0411140.

\bibitem{navarrosantiago}
I.~Navarro and J.~Santiago,
``Gravity on codimension 2 brane worlds,''
arXiv:hep-th/0411250.

\bibitem{Redi}
M.~Redi,
``Footballs, Conical Singularities and the Liouville Equation,''
arXiv:hep-th/0412189.

\bibitem{Kofinas:2004ae}
G.~Kofinas,
``Conservation equation on braneworlds in six dimensions,''
arXiv:hep-th/0412299.

\bibitem{Rubakov:1983bz}
V.~A.~Rubakov and M.~Shaposhnikov,
``Extra Space-Time Dimensions: Towards A Solution To The Cosmological Constant Problem,''
Phys.\ Lett.\ B {\bf 125}, 139 (1983).

\bibitem{Randjbar-Daemi:1985wg}
S.~Randjbar-Daemi and C.~Wetterich,
``Kaluza-Klein Solutions With Noncompact Internal Spaces,''
Phys.\ Lett.\ B {\bf 166}, 65 (1986).

\bibitem{Wetterich:1984rv}
C.~Wetterich,
``The Cosmological Constant And Noncompact Internal Spaces In Kaluza-Klein Theories,''
Nucl.\ Phys.\ B {\bf 255}, 480 (1985).

\bibitem{Wetterich:wd}
C.~Wetterich,
``Kaluza-Klein Cosmology And The Inflationary Universe,''
Nucl.\ Phys.\ B {\bf 252}, 309 (1985).

\bibitem{Gibbons:1986wg}
G.~W.~Gibbons and D.~L.~Wiltshire,
``Space-Time As A Membrane In Higher Dimensions,''
Nucl.\ Phys.\ B {\bf 287}, 717 (1987)
[arXiv:hep-th/0109093].

\end{thebibliography}
\end{document}